\documentclass[aps,pra,reprint,amsmath,amssymb,nofootinbib]{revtex4-1}

\usepackage{natbib}
\setcitestyle{authoryear,round}
\usepackage[usenames, dvipsnames]{color}
\usepackage[colorlinks,citecolor=BrickRed,urlcolor=BrickRed,linkcolor=BrickRed,hyperindex,breaklinks]{hyperref}

\usepackage{amsthm,units}

\newcommand{\ket}[1]{\ensuremath{|{#1\rangle}}} 
\newcommand{\bra}[1]{\ensuremath{{\langle #1}|}}
\newcommand{\braket}[2]{\ensuremath{{\langle #1}|{#2 \rangle}}}

\newcommand{\op}[1]{\hat{#1}}
\newcommand{\D}{\text{d}}
\newcommand{\I}{\text{i}}
\newcommand{\E}{\text{e}}

\begin{document}

\title{Entanglement, scaling, and the meaning of the\\ wave function in protective measurement}

\author{Maximilian Schlosshauer and Tangereen V.\ B.\ Claringbold}

\affiliation{\small Department of Physics, University of Portland, 5000 North Willamette Boulevard, Portland, Oregon 97203, USA}

\begin{abstract}
We examine the entanglement and state disturbance arising in a protective measurement and argue that these inescapable effects doom the claim that protective measurement establishes the reality of the wave function. An additional challenge to this claim results from the exponential number of protective measurements required to reconstruct multi-qubit states. We suggest that the failure of protective measurement to settle the question of the meaning of the wave function is entirely expected, for protective measurement is but an application of the standard quantum formalism, and none of the hard foundational questions can ever be settled in this way.\\[-.2cm]

\noindent To appear in: \emph{Protective Measurements and Quantum Reality: Toward a New Understanding of Quantum Mechanics}, edited by S.\ Gao (Cambridge University Press, Cambridge, 2014).
\end{abstract}

\maketitle

From the start, the technical result of protective measurement has been suggested to have implications for the interpretation of quantum mechanics. Consider how \citet{Aharonov:1993:qa} chose to begin their original paper introducing the idea of protective measurement:
\begin{quote}
We show that it is possible to measure the Schr\"odinger wave of a single quantum system. This provides a strong argument for associating physical reality with the quantum state of a single system \dots\,.
\end{quote}
Since then, the pioneers of protective measurement seem to have taken a more moderate stance. \citet{Vaidman:2009:po}, in a recent synopsis of protective measurement, concedes that
\begin{quote} 
the protective measurement procedure is not a proof that we should adopt one interpretation instead of the other, but it is a good testbed which shows advantages and disadvantages of various interpretations.
\end{quote}
Notwithstanding this more subtle perspective and a number of critical studies of the technical and foundational aspects of protective measurement,\footnote{See, for example, \citet{Schwinger:1993:im,Rovelli:1994:ll,Samuel:1994:um,Unruh:1994:ll,Dass:1999:az,Alter:1997:oo,Uffink:1999:zz,Uffink:2013:st}.} \citet{Gao:2013:om} has maintained, if not amplified, the force of Aharonov and Vaidman's original argument:
\begin{quote}
An immediate implication is that the result of a protective measurement, namely the expectation value of the measured observable in the measured state, reflects the actual physical property of the measured system, as the system is not disturbed after this result has been obtained. \dots\ Moreover, since the wave function can be reconstructed from the expectation values of a sufficient number of observables, the wave function of a quantum system is a representation of the physical state (or ontic state) of the system.
\end{quote}
Clearly, if we could reliably measure the unknown quantum state of a single quantum system without changing that state, it would be entirely sensible---and perhaps even inevitable---to admit the objective, physical reality of this state. Such a measurement, however, is impossible, and no measurement scheme based on an application of the standard quantum formalism, protective measurement included, can rise above this intrinsic limitation \citep{Ariano:1996:om}.

It follows that whatever form the ``measurement of the wave function'' takes in protective measurement, it must be weaker than the condition we stated in the previous paragraph: ``If we could \emph{reliably} measure the \emph{unknown} quantum state of a single quantum system \emph{without changing} that state \dots\,.'' The italicized words indicate possibilities for relaxing this condition. We might be content with measurements that are not 100\% reliable and may change the state, as long as the disturbance can be made arbitrarily small (or unlikely). Or we might be able to show that the measurement is possible only for certain quantum states, or under certain conditions, or both. 

Indeed all of these concessions must be made in the case of protective measurement (\citealp{Aharonov:1993:jm,Dass:1999:az}; see also \citealp{Uffink:1999:zz,Uffink:2013:st}). Of these, the fact that protective measurement only works under carefully designed conditions and for special quantum states---specifically, the system must be in a nondegenerate eigenstate of its Hamiltonian---may well be of least concern. After all, if protective measurement allowed us to operationally establish the reality of an unknown quantum state in certain situations, perhaps it would not be so far-fetched to extend this interpretation to the rest of the states. The more serious issue, however, arises from the inevitable system--apparatus entanglement in protective measurement. This entanglement introduces an irreducible randomness into the readout; there is a nonzero probability for the system to end up in a state different from the initial state. While this issue has been pointed out before \citep{Alter:1997:oo,Dass:1999:az}, here we will take it up in more detail, by describing the creation of entanglement in protective measurement (Sec.~\ref{sec:theory-entangl-prot}) and discussing the implications for the claim that protective measurement suggests the reality of the wave function (Sec.~\ref{sec:impl-entangl-prot}). In Sec.~\ref{sec:prot-meas-scal}, we will identify another, more subtle challenge to this claim, namely, the exponential scaling of the number of protective measurements required to reconstruct multi-qubit states. We end on a broader note (Sec.~\ref{sec:prot-meas-quant}), arguing that since protective measurement is an application of the quantum formalism, it cannot settle significant foundational questions.

\section{\label{sec:theory-entangl-prot}Theory of entanglement in protective measurement}

Here we will go beyond the zeroth-order limit $T\rightarrow\infty$ usually considered in protective measurement \citep{Aharonov:1993:qa,Aharonov:1993:jm,Dass:1999:az} and derive an expression for the final system--apparatus state to first order in $1/T$. 

Consider two systems $S$ and $A$ described by self-Hamiltonians $H_S$ and $H_A$, respectively. System $A$ plays the role of a measuring apparatus for $S$, in the sense that we let $S$ and $A$ interact such that information about $S$ can be transferred to $A$. The interaction is generated by the interaction Hamiltonian
\begin{equation}\label{eq:lalaa}
\op{H}_I(t) = g(t) \op{P} \otimes \op{O},
\end{equation}
where $\op{P}$ is the momentum operator of $A$ (the canonical conjugate to the position operator $\op{X}$), and $\op{O}$ is an arbitrary observable of the system $S$. The function $g(t)$, with $\int_0^T g(t)\,\D t=1$, describes the time-dependent strength of the interaction, with $g(t)=0$ for $t<0$ and $t>T$. Thus, $T$ describes the duration of the measurement interaction. In contrast with a standard impulsive (strong, von Neumann) measurement, here $T$ is taken to be very large. The normalization condition $\int_0^T g(t)\,\D t=1$ then implies that the magnitude of $g(t)$ will be small. This results in a weak adiabatic coupling between the system and the apparatus. Neglecting the switching-on and switching-off periods around $t=0$ and $t=T$ and assuming $g(t)$ remains approximately constant for $t \in [0,T]$, we can write $g(t)= 1/T$. Thus, the total Hamiltonian can be treated as time-independent for the duration of the measurement interaction,
\begin{equation}\label{eq:ham}
\op{H}= \op{H}_S+\op{H}_A+\frac{1}{T}\left(\op{P} \otimes \op{O}\right).
\end{equation}
To simplify the formal treatment from here on, let us assume that $\op{H}_A$ commutes with $\op{P}$.\footnote{This assumption is not necessary for a protective measurement to obtain \citep{Dass:1999:az}.} Then we can find a set of simultaneous eigenstates $\{\ket{A_i}\}$ of $\op{H}_A$ and $\op{P}$ such that
\begin{equation}\label{eq:haaz}
\op{H}_A\ket{A_i}=E_i^A\ket{A_i}, \qquad \op{P}\ket{A_i}=a_i\ket{A_i}.
\end{equation}
Therefore, the exact eigenstates of the full Hamiltonian $\op{H}$ can be written in product form as $\ket{E_m^S(a_i)}\ket{A_i}$, where the $\ket{E_m^S(a_i)}$ are the eigenstates of the system part of $\op{H}$, which is given by
\begin{equation}
\op{H}'_S(a_i) = \op{H}_S+\frac{1}{T} a_i \op{O}.
\end{equation}
Note that $\op{H}'_S(a_i)$, and thus also its eigenstates $\ket{E_m^S(a_i)}$, explicitly depend on the eigenvalue $a_i$ of $\op{P}$.

Suppose now that $S$ is initially in a nondegenerate eigenstate $\ket{n}$ of $\op{H}_S$ (but not necessarily of $\op{O}$\footnote{See the discussion by \citet{Uffink:1999:zz,Uffink:2013:st} and \citet{Gao:2013:st}.}) with eigenvalue $E_n$, and let the pointer of $A$ be described by a Gaussian wavepacket $\ket{\phi(x_0)}$ of eigenstates of $\op{X}$ centered around $x_0$. Thus, the initial composite state of system and apparatus is
\begin{equation}
\ket{\Psi(t=0)} = \ket{n}\ket{\phi(x_0)}.
\end{equation}
Since $\op{H}$ is time-independent, at $t=T$ this state has evolved into (taking $\hbar\equiv 1$)
\begin{equation}
\ket{\Psi(t=T)} = \E^{-\I \op{H} T} \ket{n}\ket{\phi(x_0)}.
\end{equation}
Inserting a complete set of eigenstates $\ket{E_m^S(a_i)}\ket{A_i}$ of $H$, we obtain
\begin{align}
&\ket{\Psi(t=T)} =\nonumber\\ &= \E^{-\I \op{H} T} \left(\sum_{m,i} \ket{E_m^S(a_i)}\ket{A_i}\bra{A_i}\bra{E_m^S(a_i)}\right) \ket{n} \ket{\phi(x_0)} \nonumber\\
&= \E^{-\I \op{H} T} \sum_{m,i} \left(\braket{E_m^S(a_i)}{n} \braket{A_i}{\phi(x_0)}\right) \ket{E_m^S(a_i)}\ket{A_i} \nonumber\\
&= \sum_{m,i} \E^{-\I E(m,a_i) T} \left(\braket{E_m^S(a_i)}{n} \braket{A_i}{\phi(x_0)}\right) \ket{E_m^S(a_i)}\ket{A_i},\label{eq:jkbknjd}
\end{align}
where
\begin{align}\label{eq:isvgaa}
E(m,a_i) &= E^A_i + \frac{1}{T}a_i \bra{E_m^S(a_i)}\op{O} \ket{E_m^S(a_i)} \nonumber \\ &\quad + \bra{E_m^S(a_i)} \op{H}_S \ket{E_m^S(a_i)}
\end{align}
are the eigenvalues of $\op{H}$ corresponding to the states $\ket{E_m^S(a_i)}\ket{A_i}$.

By regarding $\op{H}_I$ as a perturbation to $\op{H}_S+\op{H}_A$, we can write down the perturbative expansion of the exact eigenstates $\ket{E_m^S(a_i)}\ket{A_i}$ of $\op{H}$,
\begin{multline}
\ket{E_m^S(a_i)}\ket{A_i} = \ket{m}\ket{A_i} \\ + \frac{1}{T}\left( \sum_{m'\not= m} \frac{\bra{m'}\op{O}\ket{m}}{E_m-E_{m'}}\ket{m'}\right) a_{i}\ket{A_i}  + \mathcal{O}(1/T^2).\label{eq:isvgaa}
\end{multline}
In the limit $T\rightarrow\infty$ usually considered in the treatment of protective measurement, we can therefore replace all states $\ket{E_m^S(a_i)}$ in Eq.~\eqref{eq:jkbknjd} by the unperturbed eigenstates $\ket{m}$ of $\op{H}_S$. Reintroducing the operators $\op{H}_A$ and $\op{P}$ in the exponent of the time-evolution operator, this results in the nonentangled final state 
\begin{align}
&\ket{\Psi(t=T)} =\nonumber\\ &= \sum_{i} \E^{-\I \op{H}_A T -\I E_nT -\I \op{P} \bra{n}\op{O} \ket{n}} \braket{A_i}{\phi(x_0)} \ket{n}\ket{A_i} \nonumber \\
&= \E^{-\I E_nT} \ket{n}\E^{-\I \op{H}_A T} \E^{-\I \op{P} \bra{n}\op{O} \ket{n}}\sum_{i} \braket{A_i}{\phi(x_0)} \ket{n}\ket{A_i} \nonumber \\
&= \E^{-\I E_nT} \ket{n} \E^{-\I \op{H}_A T}\E^{-\I \op{P} \bra{n}\op{O} \ket{n}} \ket{\phi(x_0)}.
\end{align}
Since $\E^{-\I \op{P} \Delta x}$ is the translation operator, the term $\E^{-\I \op{P} \bra{n}\op{O} \ket{n}}$ applied to the initial wave packet $\ket{\phi(x_0)}$ will shift the center of the wave packet by an amount equal to $\bra{n}\op{O} \ket{n}\equiv \langle \op{O} \rangle_n$, which is the expectation value of $\op{O}$ in the initial state $\ket{n}$ of the system. Thus, to zeroth order, the final system--apparatus state is
\begin{align}\label{eq:gctctwg}
\ket{\Psi(t=T)} = \E^{-\I E_nT} \ket{n} \E^{-\I \op{H}_A T}\ket{\phi(x_0+\langle \op{O} \rangle_n)}.
\end{align}
This establishes the familiar main result of protective measurement: information about the expectation value of $\op{O}$ in state $\ket{n}$ has been transferred to the apparatus \citep{Aharonov:1993:qa,Aharonov:1993:jm,Dass:1999:az}. 

The crucial point, however, for our subsequent discussion is the observation that for any finite value of $T$, the system--apparatus state \eqref{eq:jkbknjd} is entangled, and therefore the initial state of the system, $\ket{n}$, has been changed. To explicitly see this, we insert expansion \eqref{eq:isvgaa} into Eq.~\eqref{eq:jkbknjd}. Keeping only terms up to $\mathcal{O}(1/T)$ and using the first-order perturbative approximation to the energy eigenvalues $E(m,a_i)$, $E(m,a_i) \approx E^A_i + \frac{1}{T}a_i \langle\op{O} \rangle_m + E_m$, we find (again reintroducing the operators $\op{H}_A$ and $\op{P}$)
\begin{multline}
\ket{\Psi(t=T)} = \E^{-\I E_nT} \ket{n} \E^{-\I \op{H}_A T}\ket{\phi(x_0+\langle \op{O} \rangle_n)} \\
+ \frac{1}{T} \E^{-\I \op{H}_A T}\sum_{m\not= n} \frac{\bra{m}\op{O}\ket{n}}{E_n-E_m} \bigl[ \E^{-\I E_nT} \E^{-\I \op{P} \langle \op{O} \rangle_n} \\- \E^{-\I E_mT} \E^{-\I \op{P} \langle \op{O} \rangle_m}\bigr]\ket{m} \ket{\widetilde{\phi}(x_0)},
\end{multline}
where $\ket{\widetilde{\phi}(x_0)} = \sum_i a_i \ket{A_i} \braket{A_i}{\phi(x_0)}$ is a distorted version of the initial pointer wave packet $\ket{\phi(x_0)}=\sum_i \ket{A_i} \braket{A_i}{\phi(x_0)}$. (One may also include the second-order perturbative correction to the energy eigenvalues such that the argument of the time-evolution operator is to first order in $1/T$; this correction, however, is irrelevant to the argument below and will therefore be neglected.) The operator $\E^{-\I \op{P} \langle O\rangle}$ then shifts $\ket{\widetilde{\phi}(x_0)}$ by an amount $\langle \op{O} \rangle$, leading to the final state
\begin{multline}
\ket{\Psi(t=T)} = \E^{-\I E_nT} \ket{n} \E^{-\I \op{H}_A T}\ket{\phi(x_0+\langle \op{O} \rangle_n)} 
\\+ \frac{1}{T} \E^{-\I \op{H}_A T} \sum_{m\not= n} \frac{\bra{m}\op{O}\ket{n}}{E_n-E_m}\ket{m} \bigl[ \E^{-\I E_nT} \ket{\widetilde{\phi}(x_0+ \langle \op{O} \rangle_n)} \\-  \E^{-\I E_mT} \ket{\widetilde{\phi}(x_0+ \langle \op{O} \rangle_m)}\bigr].
\label{eq:csgs}
\end{multline}
The first term on the right-hand side is the familiar zeroth-order term of Eq.~\eqref{eq:gctctwg}. The second term represents quantum correlations between all other eigenstates $\ket{m}\not=\ket{n}$ of $\op{H}_S$ and wave packets representing shifted apparatus pointers. Thus, Eq.~\eqref{eq:csgs} describes an entangled superposition involving all possible energy eigenstates $\{\ket{m}\}$ of the system correlated with pointer states. In particular, we see that a readout of the apparatus pointer will indicate, with probability proportional to $1/T^2$, the expectation value of $\op{O}$ in a state $\ket{m}$ orthogonal to the initial state $\ket{n}$ of the system, with the system then left in this orthogonal state $\ket{m}$ and not in the initial state $\ket{n}$. We also see that, again with probability proportional to $1/T^2$, the apparatus pointer may indicate the expectation value of $\op{O}$ in the initial state $\ket{n}$ while the system has been projected onto the orthogonal subspace spanned by $\{\ket{m}\}_{m\not=n}$.

\section{\label{sec:impl-entangl-prot}Implications of entanglement in protective measurement}

The finite system--apparatus entanglement arising in a protective measurement entails that protective measurements can never transcend the irreducibly probabilistic element inherent in any measurement of a (fully or partially) unknown quantum state. The problem is not only that the quantum state reconstructed from protective measurements may well be (unpredictably) different from the initial state of the system we had set out to measure. The problem is also that even on those occasions when the protective measurement succeeds---i.e., when the collapsed state of the apparatus pointer indicates the expectation value corresponding to the initial state of the system---we cannot infer from the readout of the pointer that we have indeed obtained information about the initial state of the system, rather than about any other state. This is so because \emph{there is no possibility of knowing whether we have succeeded}: while the final pointer measurement may project the system back onto its initial state, the readout itself cannot tell us whether this has actually happened. 

\citet{Gao:2013:om} misses this important point when he suggests that ``when the measurement obtains the expectation value of the measured observable, the state of the measured system is not disturbed.'' But there is a crucial difference between a situation in which the system remains in the initial state throughout, and a situation in which the system becomes entangled with the apparatus and, through a secondary measurement and with probability less than one, is subsequently projected back onto the initial state. Only the former situation would permit conclusions about having measured the initial state of the system, while it is the latter situation that applies to protective measurements. Thus, \emph{pace} Gao, the state of the measured system is \emph{always} disturbed in a protective measurement. 

This undermines the claim that the reconstructed quantum state must be a real, objective property, in the sense that it must have already existed prior to the measurement \citep[for a similar conclusion, see][]{Alter:1996:oo,Alter:1997:oo,Dass:1999:az}. Indeed any measurement worth its name---any measurement that allows us to obtain \emph{new} information about the system---will entangle the system and the apparatus, thus introducing an element of randomness and, in this sense, disturbing the initial state of the system; this is as true for a protective measurement as it is for any other kind of quantum measurement. Whether the measurement is strong, weak, protective, or ``reversible'' \citep{Ueda:1992:zz,Imamoglu:1993:gb}; whether we perform a sequence of measurements on a single system or just one measurement: the maximum possible information gain will always be the same \citep{Ariano:1996:om}.\footnote{\citet{Alter:1996:oo} have constructed a scheme for measuring a single squeezed harmonic-oscillator state in such a way that the final system--apparatus state is deterministically returned to a disentangled state. But as the authors themselves point out, implementation of this scheme requires full \emph{a priori} knowledge of the state of the system, which means that no information is gained in such a ``measurement'' \citep[see also][]{Alter:1995:oo,Alter:1997:oo}.} 

Elaborating on his claim that protective measurement shows that ``the wave function of a quantum system is a representation of the physical state of the system,'' \citet{Gao:2013:om} dismisses concerns about state disturbance by arguing that the probability for collapsing the system's state to an orthogonal outcome ``can be made arbitrarily small in principle when $T$ approaches infinity.'' But, from the foundational point of view relevant here, \emph{any} nonzero value of this probability, no matter how small it may be made, will spoil the claim that protective measurement permits us to learn the initial state of the system and that it thus demonstrates the reality of the wave function. The limit $T \rightarrow \infty$ required for a reliable protective measurement can never be attained---and if it \emph{could}, we would only have time, so to speak, to protectively measure one observable, thus precluding the possibility of reconstructing the wave function. It is impossible \emph{in principle} to reliably determine the expectation values required to reconstruct the wave function; it is not just impossible in practice. (We agree with \citealt{Englert:2013:uu} that to call something ``possible in principle'' is simply meaningless. Either, whatever action is contemplated is \emph{possible in practice}, or it is \emph{impossible in principle}.)

The issue at stake here bears some similarity to the situation in quantum tomography, where a quantum state is reconstructed from projective measurements on an ensemble of identically prepared systems. Of course, in the hypothetical case of infinitely many measurements on an infinitely large ensemble, the state determination would be exact. But it is impossible in principle to generate such an ensemble, or to carry out infinitely many measurements. Consider one more related, and this time purely classical, example. It is well known that the concept of probability cannot be derived from relative frequencies, simply because we cannot have an infinite number of trials. For any finite number of trials, all we can say is that is it \emph{unlikely} that the measured relative frequency will deviate much from the probability we have inferred from that frequency. ``Unlikely,'' however, already presumes a notion of probability, making the derivation circular. To reply to this conclusion by saying that one may consider infinitely many trials or infinitely large ensembles ``in principle'' is to miss the point: we are, after all, aiming at a fundamental definition that connects an abstract concept (probability) with what can be measured (relative frequencies). Infinitely large ensembles simply do not exist, and no matter how large we may make the ensemble, at the end we must make a quantitative, \emph{probabilistic} judgment about the correspondence between the theoretical probability value and the relative frequency value. 

The case of protective measurement highlights how vigilant one needs to be when using results obtained from mathematical idealizations to justify conclusions pertaining to fundamental questions of nature and interpretation. Just as in quantum tomography, lack of accuracy may not matter as far as practical implementation is concerned: a method that gives us an approximate picture is often all we need. But in-principle lack of accuracy may become decisive when the very procedure is claimed to have implications for our conceptual understanding of the theory itself. It certainly is decisive in the case of protective measurement, refuting the claim that protective measurement has demonstrated the ontological status of the wave function. It follows that secondary claims based on this claim must fail, too; an example is Gao's \citeyearpar{Gao:2013:ga} suggestion that protective measurement effortlessly establishes a result equivalent to the theorem derived by \citet{Pusey:2012:np}---namely, that the wave function must be ``uniquely determined by the underlying physical state.''\footnote{It should be mentioned that such a result does not conclusively follow from the theorem of \citet{Pusey:2012:np} either; for critical discussions, see \citet{Colbeck:2012:cr,Hardy:2012:rr,Schlosshauer:2012:pr,Schlosshauer:2014:pr,Walden:2013:st}.}

\section{\label{sec:prot-meas-scal}The scaling problem}

We now turn our attention to the question of the number of protective measurements of expectation values required to (approximately) reconstruct a wave function; in particular, we will analyze how this number scales with the size of the system. This, at first glance, may appear to be a question of purely practical concern. However, as we will indicate below, it may have foundational implications as well.

Consider a qubit described by an arbitrary density matrix $\op{\rho}$. In the Bloch representation, this density matrix can be written as
\begin{equation}
\op{\rho}=\frac{1}{2}\left(\mathbb{I}+\mathbf{n}\cdot \ensuremath{\mathbf{\op{\boldsymbol \sigma}}}\right).
\end{equation}
Here, $\mathbb{I}$ denotes the identity operator. The components of the operator $\ensuremath{\mathbf{\op{\boldsymbol \sigma}}}$ are given by the Pauli operators $\op{\sigma}_x$, $\op{\sigma}_y$, and $\op{\sigma}_z$, and the real-valued components of the vector $\mathbf{n}$ are given by the expectation values of $\op{\sigma}_x$, $\op{\sigma}_y$, and $\op{\sigma}_z$ in the state $\op{\rho}$, i.e., 
\begin{equation}
n_i = \text{Tr}\left[\op{\sigma}_i\op{\rho}\right]=\left\langle \op{\sigma}_i \right\rangle_{\op{\rho}}, \qquad i=x,y,z.
\end{equation}
It follows that the density matrix of a qubit is uniquely determined by the three expectation values $\left\langle \op{\sigma}_x \right\rangle_{\op{\rho}}$, $\left\langle \op{\sigma}_y \right\rangle_{\op{\rho}}$, and $\left\langle \op{\sigma}_z \right\rangle_{\op{\rho}}$. We are free, of course, to choose other triples of observables, as long as these observables form an informationally complete set (i.e., as long as their expectation values uniquely determine the qubit state). But we always need to measure the expectation values of three such observables to reconstruct an arbitary state $\op{\rho}$; no pair of observables will do. 

In the general case of an $d$-dimensional Hilbert space, the measurement of the expectation values of a minimum of $n(d)=d^2-1$ observables will be required to determine an arbitrary state $\op{\rho}$. Therefore, determining an arbitrary $N$-qubit state requires at least 
\begin{equation} 
n(N)=d^2-1 = (2^N)^2 -1 = 4^N-1
\end{equation}
expectation values to be measured, which means that the required number of measurements grows exponentially with $N$. The number of observables can be reduced, however, if prior information about the state $\op{\rho}$ is available. For example, if $\op{\rho}$ is known to be pure, then the number of observables required to uniquely determine the state only scales linearly with $d$ \citep{Heinosaari:2013:az,Chen:2013:oo}. Protective measurement may avail itself to such a reduction in the number of required measurements, since the system has to be in a pure eigenstate of the self-Hamiltonian of the system. Even so, since $d=2^N$ for an $N$-qubit state, this still results in an exponential scaling behavior.

Since no concrete experimental realizations of protective measurement are presently available, it is difficult to provide a good estimate of the time that would be required to reconstruct the wave function of a multi-qubit system with a degree of accuracy comparable to that typically achieved in quantum tomography. \citet{Dickson:1995:lm} points out that the interaction time $T$ may only need to be large on an atomic scale; without giving further details, he provides an estimate of $10^{-5}$ seconds. But this value seems unduly low in light of the fact that even the impulsive measurements used in quantum tomography often take longer. For example, \citet{Haffner:2005:sc} have carried out quantum tomography on an eight-qubit state of trapped $^{40}\text{Ca}^+$ ions, requiring 656,100 measurements and a total measurement time of 10 hours, or about 0.05~seconds per measurement. 

Whatever estimates for the measurement duration $T$ might be reasonable in potential practical implementations of protective measurement, it is clear that because of the exponential scaling, reconstructions of quantum states of large systems (say, hundreds of qubits) would require astronomically long total measurement times. Of course, this is no different from standard quantum tomography. But contrary to protective measurement, quantum tomography has not been associated with the claim that it demonstrates the reality of the quantum state (the reason being, we would expect the proponent of protective measurement to argue, that quantum tomography works with ensembles, not single systems). Thus, there is a much greater burden on protective measurement to show that its method for state reconstruction has the suggested physical meaning and foundational implications. 

To reconstruct a 100-qubit pure state would require the measurement of about $2^{100} \approx 10^{30}$ expectation values, and even if we accept Dickson's optimistic estimate of $T \approx \unit[10^{-5}]{s}$ for a single protective measurement, the time needed to measure such a state would exceed the present age of the universe by many orders of magnitude. Is this merely a practical problem? We are not sure it is. Clearly, in contrast with the discussion in the preceding section, the issue is no longer about the tension between a strict mathematical limit and the inevitably finite version attainable in practice. Yet, even if we assume that the universe will continue to exist for a sufficiently long time for such a large number of measurements to be carried out, there is certainly no sense in which this experiment could ever be realized. As \citet{Englert:2013:uu} put it,
\begin{quote}
Statements like ``In principle, I could solve the Schr\"odinger equation to predict the next solar eclipse'' are empty unless you can do it in practice. 
\end{quote}
By the same token, statements like ``In principle, I could carry out a protective measurement of a 100-qubit state to establish its reality'' must be considered empty. But if there is no possibility for this state to be measured, protectively or otherwise, what can such a state possibly mean?\footnote{\citet{Aaronson:2007:po} has posed a similar question in the context of quantum tomography.} We must leave this question open; our aim here has been to point out that even ostensibly mundane practical constraints may have fundamental implications for the question of how and whether protective measurement could decide the question of the ontological meaning of the wave function.

\section{\label{sec:prot-meas-quant}Protective measurement and the quantum formalism}

Protective measurement does not demonstrate the physical reality of the wave function. Should this result be surprising? Did we, and other authors \citep{Schwinger:1993:im,Rovelli:1994:ll,Samuel:1994:um,Unruh:1994:ll,Dass:1999:az,Alter:1997:oo,Uffink:1999:zz,Uffink:2013:st}, really need to invoke various, ostensibly technical arguments to come to this conclusion? 

Quantum mechanics provides a formalism for relating and transforming probability assignments concerning outcomes of future measurements. The notion of measurement and the existence of outcomes are all taken to be primitives of the theory. (To repeat a popular analogy, this is just as in classical probability theory, which neither explains the existence of dice and nor why throwing them results in particular results.) To want to say more is to tack onto the quantum formalism a story: of metaphysics, say, or of causation. But as the wealth of competing interpretations of quantum mechanics shows, the choice of any such particular story is hopelessly underdetermined by the quantum formalism itself. The formalism does not mandate any particular ontological commitment toward the interpretation of its elements, quantum states and their corresponding probabilities included.\footnote{Indeed, one can construct a picture of quantum mechanics in which quantum states are nothing but a representation of our personal beliefs about our future experiences when we interact with a quantum system \citep{Fuchs:2013:aa}.} When we calculate probabilities from quantum states, we start from some initial quantum-state assignment. But the question of what this assignment physically \emph{means} or \emph{represents} is of no relevance, because any probabilistic predictions derived from the quantum formalism, using the initial state assignment, are insensitive to how we choose to answer the question. We get the quantum formalism cranking to obtain a new quantum state, and there is no reason to apply to this state an interpretation different from the interpretation we chose to give our original state assignment. 

The point here is that no application of the quantum formalism, and no observational data that is in agreement with the predictions of this formalism, can provide definite answers to questions about the interpretation of the quantum state. Protective measurement is just such an application. Therefore, it is not equipped to settle the significant foundational and interpretive questions, no matter how wishful the thinking. (By ``significant'' we mean the hard questions---the question of the meaning of the wave function, for example---rather than the ``softer'' questions about the explanatory power or the reasonableness of individual interpretations of quantum mechanics.)

Of course this is not to say that by milking the quantum formalism we cannot produce something fresh. Quantum information theory and decoherence theory are good examples, but they, just like protective measurement, have not answered the hard interpretive questions; and they, too, could not be expected to do so. Quantum information theory may have motivated new information-based interpretations of quantum mechanics, but there are quantum information theorists who are Bohmians and others who are Everettians. Decoherence, it is to be remembered, is an essentially technical result about the dynamics and measurement statistics of open quantum systems. In particular, its predictively relevant part relies on reduced density matrices, whose formalism and interpretation presume the collapse postulate and Born's rule. Thus if we understand the quantum measurement problem as the question of how to reconcile the linear, deterministic evolution described by the Schr\"odinger equation with the occurrence of random, definite measurement outcomes, then decoherence has certainly not solved this problem, as is now widely recognized \citep{Schlosshauer:2003:tv,Schlosshauer:2007:un}. What decoherence rather solves is a \emph{consistency problem}: the problem of explaining why and when quantum probability distributions approach the classically expected distributions. But this is a purely practical problem, not a game-changer for quantum foundations. To be sure, the picture associated with the decoherence process has sometimes been claimed to be suggestive of particular interpretations of quantum mechanics\footnote{Indeed, historically decoherence theory arose in the context of Zeh's independent formulation of an Everett-style interpretation \citep{Zeh:1970:yt,Camilleri:2009:aq}.} or to pinpoint internal concistency issues \citep{Schlosshauer:2003:tv}. But it might be safer to say that certain interpretations (such as the Everett interpretation) are simply more \emph{in need} of decoherence to define their structure. At the end of the day, any interpretation that does not involve entities, claims, or structures in contradiction with the prediction of decoherence theory (which is to say, with the predictions of quantum mechanics) will remain viable.

It follows that if we hope to make headway in foundational matters, we have to consider theories beyond quantum mechanics and study how their predictions match those of quantum mechanics. Reconstructions of quantum mechanics are one example of this approach; they have shown that features traditionally regarded as uniquely quantum---such as interference, Bell-type violations, no-signaling and no-cloning constraints, and state disturbance through measurement---are generic to entire classes of probabilistic theories. Another example is Bell's theorem \citep{Bell:1964:ep,Bell:1966:ph}, although what exactly the experimentally measured violations of Bell's inequalities tell us about nature remains a matter of debate \citep{Schlosshauer:2011:ee}. Like Bell's theorem, the PBR theorem \citep{Pusey:2012:np} is based on the consideration of  hidden-variables models and accommodates a variety of conclusions \citep{Colbeck:2012:cr,Hardy:2012:rr,Schlosshauer:2012:pr,Schlosshauer:2014:pr}. Thus, a decisive answer to a foundational question may elude us even if we consider models beyond quantum mechanics.

\section{Concluding remarks}

In response to Uffink's \citeyearpar{Uffink:1999:zz} criticism of protective measurement, \citet{Gao:2011:om} writes:
\begin{quote}
It seems that the errors in Uffink's arguments were made at least partly due to his biased philosophical opinions. Why protect the interpretation of the wave function against protective measurements? Why make the different views on the meaning of the wave function peacefully coexist? Is it not very exciting and satisfying if we can decide the issue of the interpretation of the wave function someday? Is it not one of the ultimate objectives of our explorations in quantum foundations? 
\end{quote}
To this, \citet{Uffink:2012:om} replies:
\begin{quote}
Of course, [I agree fully] with Gao that such an alternative view would be much more desirable. However, apart from the hot aspirations we might all have concerning the interpretation of quantum theory, we also need the cool breeze of critical analysis before we step forward. 
\end{quote}

Uffink's attitude, like our own, is not meant to be pessimistic. It merely reflects a realistic assessment of aims and means. Protective measurement is an ingenious implementation of a quantum measurement, but a quantum measurement it nevertheless remains. As such, it simply cannot, even in principle, accurately determine the wave function of a single system. In particular, we have pointed to two problems: the necessarily finite interaction time and the astronomically large number of measurements required for bigger systems. The probabilistic, random element of any quantum measurement remains; there cannot be any information gain without disturbance. But only if perfectly reliable, nondisturbing state determination were possible would protective measurement qualify as an arbiter in the question of the nature of the wave function. As we have argued, the failure of protective measurement to accomplish this goal is not surprising, for no application of the quantum formalism can bypass the fundamental indifference of this formalism to its interpretation.

If protective measurement had indeed established the reality of the wave function (or its direct correspondence with reality), then, without doubt, we would have happily concurred with Gao's \citeyearpar{Gao:2013:om} assessment of protective measurement as a ``paradigm shift in understanding quantum mechanics.'' As it stands, however, not only do all interpretive options remain on the table, but, in our view, protective measurement also fails to nudge us one way or the other. If one does not already believe in the reality of the wave function, then what does protective measurement offer to change one's mind? Not only does protective measurement fail to challenge the epistemic view of the wave function, but it also leaves untouched all the features that make the epistemic view so attractive and powerful in the first place \citep{Spekkens:2007:um,Fuchs:2010:az,Mermin:2012:gb}. To say so is not to diminish the practical usefulness of protective measurement or to discourage its future exploration, but to recognize the fundamental limitations when using the quantum formalism to provide its own interpretation.


\end{document}